# A CONVERT compiler of REC for PDP-8[1]

Harold V. McIntosh[*]

REC (REGULAR EXPRESSION COMPILER) is a programming language of simple structure developed originally for the PDP-8 computer of the Digital Equipment, Corporation, but readily adaptable to any other general purpose computer. It has been used extensively in teaching Algebra and Numerical Analysis in the Escuela Superior de Física y Matemáticas of the Instituto Politécnico Nacional, even to the extent of programming hand calculations with the Friden electronic desk calculator. In this way programming concepts have been introduced to the students even from their very first courses. Although the Friden machine has a memory store of only four numbers, it is of a pushdown type which already makes apparent on a rather reduced scale some of the characteristics and logical structures found in larger computers and their programming languages. Moreover, the fact that the same control language, REC, is equally applicable and equally efficient over the whole range of computer facilities available to the students gives a very welcome coherence to the entire teaching program, including the course of Mathematical Logic which is devoted to the theoretical aspects of such matters.

REC; derives its appeal from the fact that computers can be regarded reasonably well as Turing Machines with very elaborate built-in shortcuts to eliminate the grotesque inefficiency of manipulating individual bits on a single linear tape. A Turing Machine itself consists of a finite-state machine acting as the control of a tape memory; finite-state machines in turn are conveniently described by regular expressions. The REC notation is simply a manner of writing regular expression, somewhat more amenable to programming the Turing Machine which they control. If one does not wish to think so strictly in terms of Turing Machines, REC expressions still provide a means of defining the flow of control in a program which is quite convenient for many applications.

Let $\Sigma$ be an alphabet, which presumably would not contain among its letters the operational signs which we shall introduce. We then define a REC expression recursively in the following manner.

    i)    $\lambda$ is a REC expression

    ii)   ( ) is a REC expression

    iii)  if $\sigma \in \Sigma \cup \{:;\circ\}$, $\sigma$ is a REC expression

    iv)  if $\alpha$ y $\beta$ are REC expression, so is $\alpha\beta$

---

[1] This paper is seminal formal definition for REC language was published in AIM-149 of MIT Artificial Intelligence Group Jan 1968 and too "Acta Mexicana de Ciencia y Tecnología" of IPN, Jan-April 1968. REC is a programming language of extremely simple structure and what it was proved that the well publicized inconvenience of programming without a goto was a myth in Sixties endings. See Knuth, in one of his famous articles [**] says: "my dream is that by 1984 we will see a consensus developing for a really good programming language (or more likely, a coherent family of languages)". Note for Ignacio Vega-Paez [**] Knuth, D.E., "Structured programming with goto statements", Computing Surveys, Vol. 6, No. 4, Dec. 1974, pp. 261-301

[*] Escuela Superior de Física y Matemáticas, Instituto Politécnico Nacional, México.

v)   if $\sigma$ is a REC expression, so is $(\alpha)$

The operational signs are used is follows. Parentheses denote a single expression, which is to be treated as a single unit. Concatenation is implied by writing expressions in sequence. Colon [:] implies iteration of all that part of the expression which precedes the colon. Semicolon [;] terminates the concatenation of a string. The large period [∘] indicates a choice between continuing to concatenate the following expression or to pass over them until the next following colon or semicolon (if any) of the same parenthesis level is reached. Moreover such a choice is always implied following every parenthesized expression.

It is to be noted that parentheses have a very technical use in REC expressions, and are more than simple signs of grouping. Thus, since concatenation is associative, it is always written in its extended form without parentheses. When it is desired to show some particular grouping, some other Symbol, such as square brackets, should be used. The non associativity of REC parenthesization may be exploited to achieve special effects or sometimes some economy or simplification of expression.

In order that we way see the correspondence between regular expressions and REC expressions, we first show how I regular expression is to be transcribed into a REC expression.

$$\Phi \to (\,)$$
$$\lambda \to \lambda$$
$$\sigma \to \sigma$$
$$\alpha\beta \to \alpha\beta$$
$$\alpha \cup \beta \to (\circ\alpha;\beta;)$$
$$\alpha^* \to (\circ\alpha:;)$$

For the convene process of writing the regular expression corresponding to a REC expression, it is more convenient to Show how to use a REC expression to construct a transition system, whose regular expression (or class of equal regular expressions) may then be deduced. The algorithm is as follows. recursively defined.

1) A REC expression is to be read from left to right during its transcription. It will he interpreted character by character, with the exception that if a parenthesized subexpression is encountered it is to he treated recursively as an entirely new expression, When its transcription is complete additional rules govern its incorporation into the main expression.
2) For each REC expression, there will be an initial state I, and a terminal state T. During the course of transcription additional states will he formed. The last of these, a role initially filled by I, will be referred to as the "last" state during any (if the construction.
3) If $\lambda$ is seen, form a new state and draw an arrow representing a spontaneous transition from the last state to this new state.
4) If $\sigma \in \Sigma$ is seen, draw an arrow labeled $\sigma$ from the last state to a new state.
5) If : is seen, draw a spontaneous transition back to the initial state I. In addition form a new state, which will thereupon become the last state formed. It will be described as the "first state following the colon" and may or may not eventually be connected to other states.
6) If ; is seen, draw a spontaneous transition to the terminal state T. In addition, form a new state which will thereupon become the last state formed. It may or may not

  eventually be connected to other states, but if so will be referred to' as "the first state following the semicolon".
7) If ∘ is seen, draw an arrow representing a spontaneous transition to the state immediately following the next colon or semicolon, if such exists; otherwise to the terminal state T. The occurrence of the character ∘ never results in the formation of a new state.
8) If a parenthesized expression is seen, apply the entire algorithm to the expression within parentheses. When this has been done, draw an arrow representing a spontaneous transition from the last state of the exterior expression to the initial state of the interior expression. The last state of the interior expression is to be connected by a spontaneous transition to the state immediately following the next colon or semicolon of the exterior expression, if such exists; otherwise to the terminal state T of the exterior expression. In other words each parenthesized expression is to be treated rather as though it were followed by ∘. Finally, the terminal state T is to be regarded as the last state, in continuing to transcribe the REC expression.
9) The last state to be written is the accepting state of the transition system.

  With experience one can shorten the transcription somewhat; for example the initial state of a subexpression can be taken directly as the last state of the outer expression without the necessity of introducing two states and a spontaneous transition.

  In Figure 1 we present the transcription of several REC expressions. Some of the examples show how the REC expressions which are immediately derived from regular expressions are handled in order to confirm the equivalence of the two notations. The remaining example shows how a moderately complex REC expression such as $\left(RP\circ;Q\circ\left(RQ\circ;:\right):W:\right)$ would be handled.

  It will be noted that the REC expressions which are derived directly from regular expressions by the prescription we have offered form a limited class among the possible REC expressions. Or course, the remaining REC expressions must be equivalent to those so derived. In part the discrepancy is due to a desire to leave the syntax of REC expressions relatively weak, even though the result is to admit a great number of expressions which would produce useless transition diagrams; for instance we do not exclude the sequence :::::.

  Our rules for transcribing a regular expression into a REC expression has been incomplete at one minor point. Regular expressions are formed from smaller expressions by the binary operations of concatenation and union, together with the unary operation of iteration. The binary operations are associative, and thus may be compounded without the use of parentheses. However parentheses have a different significance in REC expressions, and we have not told how to use any other grouping symbol. It is clear that the unparenthesized concatenation of several regular

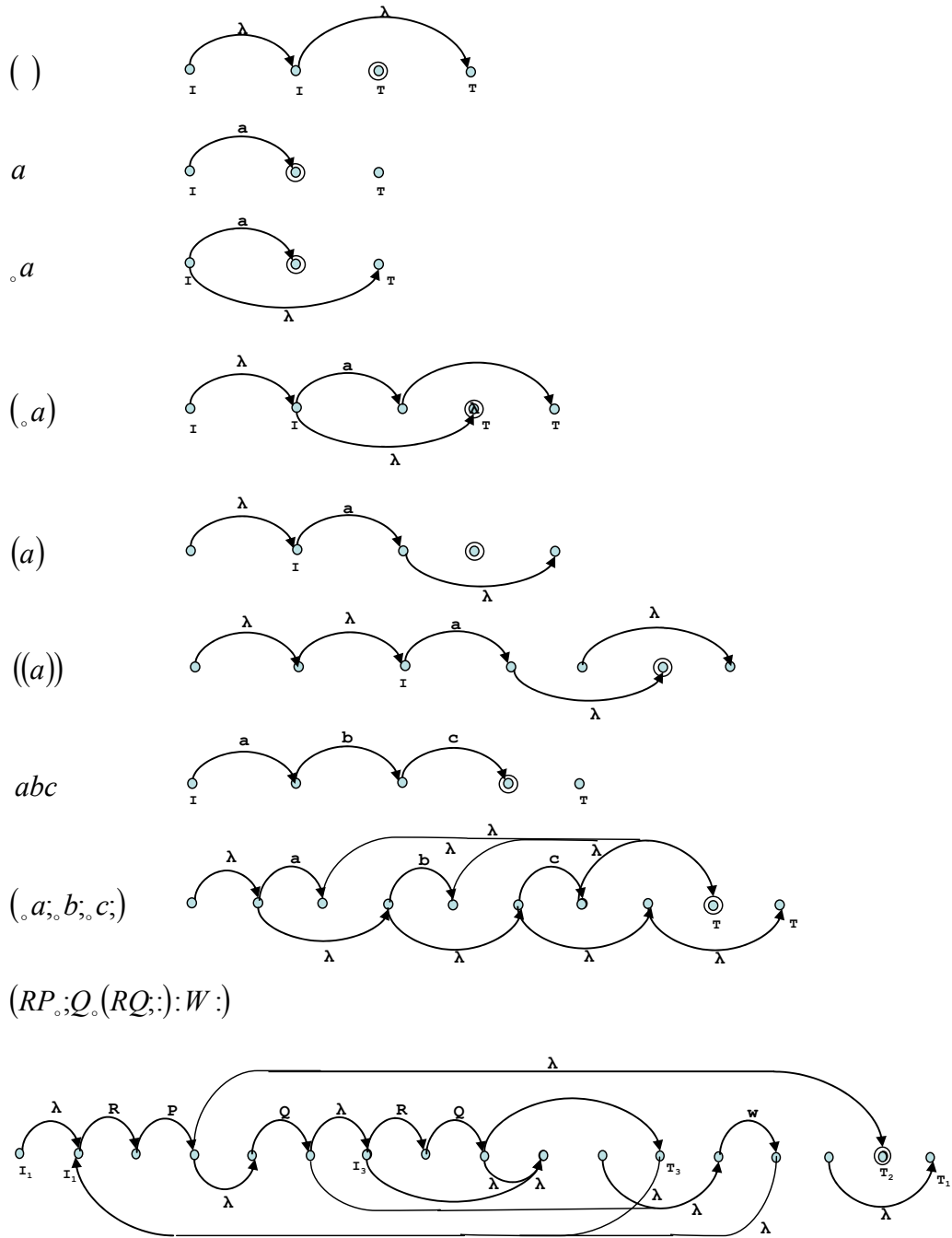

expressions passes over into the similar concatenation of the corresponding REC expressions. The associative nature of the union manifests itself when we write.
$$(AuB)uC \text{ as } (\circ(\circ A;B;);C;)$$
$$Au(BuC) \text{ as } (\circ A;(\circ B;C;);)$$
and $\qquad AuBuC \quad \text{as } (\circ A;\circ B;C;),$
all of which may be verified to produce equivalent transition systems,

To a certain extent we may think of the large period as a prefix notation for union, with the' semicolon being used to delimit its arguments. There is no point to belabor the notation, however, because the motivation of the REC notation is not merely to provide an alternative notation for writing regular expressions so much as to maintain contact with the concept of a regular expression while facilitating the writing of a class of expressions which we will have frequent occasion to Use. Thus the correspondences which we have established serve to demonstrate that the totality of REC expressions is no more nor less general than the totality of regular expressions.

Since the intention of REC expressions is to control the Operation of a general purpose computer (or more specifically a Turing Machine), we will expect the letters of the REC alphabet to represent individual operations of which the machine is capable. For this reason the letters will be called operators. Words of the REG alphabet will then correspond to sequences of operations, carried out in the order given. The transition system derived from a REC expression will then accept a word of this alphabet if it corresponds to a possible series of operations which could be carried out during the course of the calculation in question. In the case of a Turing Machine, the operators will be to write a symbol, compare a symbol to one specified, move the tape left, or move the tape right. But generally the operators will have to be chosen according to the circumstance, or the capabilities of the machine to be controlled.

In reality we are not so much interested in recognizing a possible calculation as in prescribing the particular one we want among all those possible. It is for this reason that the large period [∘] was introduced, which is related to the operation of union in a regular expression. At each place in a REC expression where ∘ occurs, there is a spontaneous transition in the transition diagram, indicating the possibility of a selection among two alternatives; to continue the regular sequence, or to start a new one by following the spontaneous transition. To specify a particular word among all those represented by a given REC expression, it is only necessary to specify this choice at each place where it becomes possible. We might even assume that there are special operators whose purpose is to make this choice. They are called *predicates*, and will always combine the symbol ∘ implicitly. Thus a predicate is combination of an appropriate operator followed by the symbol ∘ We will moreover say that a predicate takes the value *true* or *false* according to whether the decision is made to continue in the regular sequence or to follow the spontaneous transition past the nearest colon or semicolon. Every parenthesized expression is automatically assumed to be a predicate, although analysis may show that it is only capable of assuming one of the two possible values. Such was the case in the example of Figure 1.

The transition diagrams of REC expressions have two final states to accommodate their usage as predicates. Thus a calculation definitely fails, definitely succeeds, or else is in progress. Moreover the REC notation has been particularly chosen to facilitate the formation of Boolean combinations of its subexpressions. The combination AND of the predicates, *a, b, c, ...,n* is written

*(abc.. n;),*

a notation which is valid for any number of arguments. Thus always is a true predicate, whilst *a=(a;.)*.

The combination OR of these same predicates would he written
*(a; b; c;...; n;),*
which again holds for any number of arguments. *()* is a predicate which is always false, and is before, *(a;)=a*.

The complement of the predicate *x* is written
*(x).*
We accordingly always have *x=((x))*

A typical REC expression will begin with a series of operators, followed by a predicate which will decide typical questions such as whether the calculation is finished and be followed by or whether to repeat the whole procedure and be followed by ;. When these conditions fail, there will follow further calculation, expressed by a series of operators, and yet another predicate. One executes as much of a string as he can until lie meets a delimiter, and as many strings as necessary to meet a terminal condition. One practical caution which has to be observed is that if several predicates occur in a string, and one has reached the end of the string. the AND of all these predicates is true, if one arrives beyond a colon or semicolon , indicating, the string has failed, lie only knows the AND has failed, but not which individual predicate, This requires either a new test of some of the predicates. or a more cautious rewriting of the REC expression. It is one situation in which one sometimes wishes there were a more direct control of the flow of control in a REC expression; perhaps by means of labels and "GO TO's".

To give some very simple examples of the application of REC, let us bear in mind the PDP-8 computer, which has a teletype coded for 64 ASCII characters in direct communication with the central processor. Let *R* be the operator which reads one such character, either from paper tape or punched by hand on the keyboard, and *W* be the operator which sends one such character to the teletype. The characters kept in a workspace (the accumulator, say), and we may imagine 64 operators of the type *"x* which place the character *x* in this workspace erasing the previous contents, as well is 64 predicates *=x* which test the workspace for equality to the character *x*.

The REC. expression
*(R=!;W" W:)*
will doublespace everything, which it reads, until the exclamation point is encountered and it terminates operation.

Let us say that we wish to ignore all text which occurs between two stars. An appropriate expression will be
*(R=!;=*(R=*;:) : W:)*
and again it will terminate when an exclamation point is encountered in the printing text.

By including operators for the binary conversion of decimal input and output, the arithmetic operations, and a test for negative numbers, one could formulate REC expressions for arithmetic calculations. The domain of applicability of REC depends upon its complement of operators and predicates; however at present it is only the control structure which interests us.

Although we arc describing a compiler of REC for the PDP-8, the description is applicable to the majority of machines because the compilation is made entirely in terms of Subroutine calls, except for the part which corresponds to REC'S own flow of control, which is realized for the most part by appropriate transfers.

In the PDP-8, a subroutine call is made by means of the instruction *JMS* (Jump to Subroutine). Let us suppose we have the coding configuration

    X, JMS Y

     . . .

    Y, 00*

     . . .

     JMP I Y

When the instruction *JMS Y*, located at address *X*, is executed, the address *X + 1* is stored at *Y*, and transfer is made to *Y + 1*. When the subroutine is terminated, this is done by the instruction *JMP I Y*, an indirect transfer to *Y* which is a transfer to *X + 1*, so that the original program is resumed in sequence.

Data of use to the subroutine *Y* may be located at addresses *X + 1*, *X + 2*, and so on, and may be accessed indirectly through the address stored at location *Y*. By applying the instruction *ISZ Y* (Increment and skip on zero), this data may be gathered item by item. Moreover, the subroutine Y can serve as a predicate, since an *ISZ* preceding the return jump can cause a skip to *X + 2* rather than a return to *X + 1*.

In this way, the predicate, =*x*, may be treated as a composite predicate, formed from a general subroutine *EQ*, which uses the character x as a parameter in the calling sequence. =*x* would then compile into

     *IMS EQ*

     =

     *(return false)*

     *(return true)*

Clearly, this pattern accounts for predicates with multiple parameters, including none. The false return will contain a transfer, corresponding to the spontaneous transition of the transition diagram which the REC expression defines, while for the true return there will occur further subroutine jumps corresponding to subsequent operators.

With these preliminaries we may now turn to the CONVERT program REC, listing of which we give below.

```
    DEFINE ((
(REC (LAMBDA (L) (PRINTLIST (CONVERT
(QUOTE (
    OP    PAV    (=OR= R W)
```

*For the purposes of this program, three classes of letters arc distinguished: Operators (OP), Predicates (PR) and compound predicates (CP). In each category its members are listed, and treated as PAV's by the CONVERT program.*

```
    PR    PAV    (=OR= Q)
    CP    PAV    (=OR= EQ QU)
```

```
        ))
(QUOTE (
     X (XXX)
       ))
L
(QUOTE (*0
     (PR           ((JMS PR) (JMP FA)))
```

*Predicates are compiled as a subroutine call followed by a transfer to FA. FA is the heading corresponding to the FALSE exit of the segment under compilation; this transfer is skipped over when the predicate is true.*

```
     (OP           ((JMS OP)))
```

*Operators are compiled by a simple subroutine call.*

```
     ((CP X))      ((JMS CP)  (X)   (JMP FA)))
```

*Compound Predicates are, compiled as Predicates, but their parameter is included at part of the calling sequence,*

```
     ((**)         ((JMP OF) FA)
```

*The CONVERT program is written in such a way that it does not distinguish CDR of a list from a list. However, these have to be processed differently, and are therefore distinguished by a double asterisk placed in how of a fragment which has arisen at CDR of a list. When only the double asterisk is left, the end of the list has been reached, the spontaneous transition (JMP OF) corresponding to the fact that each parenthesized REC expression is regarded at a predicate it inserted, and the heading FA Is placed, since we have now arrived at the first state outside the parenthesis to which all false exits in the last segment must proceeded.*

```
     ((** CO XXX)  ((JMP HE) FA (*SKEL* FA EXPR =GNSY=
                   (=REPT= (** XXX)))))
```

*When in examining a REC expression element by element, we arrive at a colon (CTSS DOES NOT LET US WRITE ALL CHARACTERS, AN IDIOSYNCRACY OF THE LISP INPUT ROUTINE), we write a spontaneous transition to the initial state (JMP HE.), note the false exit point of all predicates in the previous segment, and define a new false exit point for the ensuing segment. The analysis continues with the remainder of the REC expression, \*\* serving as a signal that we do not deal with a new expression.*

```
     (** SC XXX)   ((JMP TR) FA (*SKEL* FA EXPR =GNSY=
                   (=REPT= (** XXX)))))
```

*When a semicolon is encountered, a spontaneous transition is made to the TRUE final state (JMP TR), the false exit point of all predicates in the previous segment is noted, and a new false exit point is established for the ensuing segment. The analysis then proceeds with the remainder of the expression.*

```
     ((** X XXX)   ((*REPT* X) (*REPT* (** XXX))))
```

*If neither delimiter is encountered, we compile the CAR and then the CDR of the expression. CAR's and CDR's are not treated uniformly because a new initial state has to be established for each subexpression, but not for each CDR.*

```
  ((===)          (=SKEL= HE EXPR =GNSY= TR EXPR
                   =GNSY= OF EXPR FA (HE (*SKEL* FA
                   EXPR =GNSY= (=REPT= (** *SAME*)))
                   TR)))
```

*In compiling a parenthesized expression, provision must be made for the initial state, TRUE final state, and FALSE final state, all of which are defined as GENSYM's. These labels must be included at appropriate points in the compiled code.*

```
)))
))))
(PRINTLIST (LAMBDA (X) (PROG (Y) (SETQ Y X) (CLOCK ()) A
(PRINT (CAR Y))
(SETQ Y (CDR Y)) (COND ((NULL Y) (RETURN (CLOCK))))
(GO A))))
```

*PRINTLIST is an auxiliary function which allows listing the compiled program with one PDP-8 instruction per line, rather than as a compact list in the usual manner that LISP would print a result.*

```
))
```

As an example of the operation of REC we may consider the following example. (REC L) is a function whose argument is the REC expression which is to be compiled. On account of inherent limitations in the orthography of the CTSS LISP input routine, certain substitutions had to be made:

SC for `;`, CO for `:`, (EQ X) for =X, (QU X) for "X.
rec ((r (eq =) co (eq :)SC  w r q w r q w r q w))

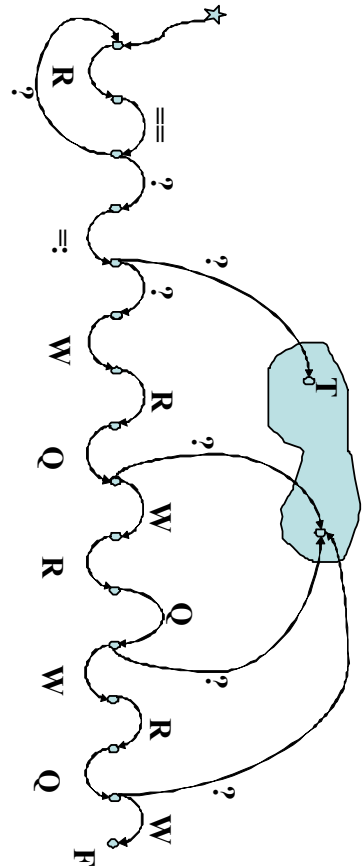

| | |
|---|---|
| GO3163 | *Initial Point* |
| (JMS R) | *R* |
| (JMS EQ) | == |
| (=) | *Parameter* |
| (JMP GO3165) | *false* |
| GO3165 | *false exit of last segment* |
| (JMS EQ) | =: |
| (:) | *parameter* |
| (JMP GO3166) | *false* |
| (JMP GO3164) | ; |
| GO3166 | *false exit of last segment* |
| (JMS W) | *W* |
| (JMS R) | *R* |
| (JMS Q) | *Q is an arbitrary predicate* |
| (JMP GO3167) | *false exit* |
| (JMS W) | *W* |
| (JMS R) | *R* |
| (JMS Q) | *Q* |
| (JMP GO3167) | *f* |
| (JMS W) | *W* |
| (JMS R) | *R* |
| (JMS Q) | *Q* |
| (JMP GO3167) | *f* |
| (JMS W) | *W* |
| (JMP FA) | *exit from last segment to FALSE final state* |
| GO3167 | *continuation on higher level, exit of F's in last segment* |
| GO3164 | *TRUE final state, exit of all semicolons* |
| 5 | *(time of execution)* |

The program which is generated is incomplete in the sense that it itself should be finished off as a subroutine, with a blank entry point bearing in appropriate label, and terminated with appropriate ISZ's and JMP I's.

For case of reference we conclude with an unannotated listing of the program.

```
     DEFINE ((
(REC (LAMBDA. (L) (PRINTLIST (CONVERT
(QUOTE (
    OP    PAV         (=OR= R W)
    PR    PAV         (=OR= Q)
    CP    PAV         (=OR= EQ QU)
    ))
(QUOTE (
    X (XXX)
    ))
L
(QUOTE (*0
    (PR              ((JMS PR) (JMP FA)))
```

```
        (OP               (JMS OP)))
        ((CP X)           ((JMS CP) (X) (JMP FA)))
        ((**)             ((JMP OF) FA))
        ((** CO XXX)      ((JMP HE) FA (*SKEL* FA EXPR =GNSY=
                          (=REPT= (** XXX)))))
        ((** SC XXX)      ((JMP TR) FA (*SKEL* FA EXPR =GNSY=
                          (=REPT= (** XXX)))))
        ((** X XXX)       ((*REPT* X) (*REPT* (** XXX))))
        ((===)            (=SKEL= HE EXPR =GNSY= TR EXPR
                          =GNSY= OF EXPR FA (HE (*SKEL* FA
                          EXPR =GNSY= (=REPT= (** *SAME*)))
                          TR)))
        )))
        ))))
(PRINTLIST (LAMBDA (X) (PROG (Y) (SETQ Y X) (CLOCK (() A
(PRINT (CAR Y))
(SETO Y (CDR Y)) (COND ((NULL Y) (RETURN (CLOCK T)))
(GO A))))
        ))
```

## ABSTRACT

REC/8 is a CONVERT program, realized in the CTSS LISP of Project MAC, for compiling REC expressions into the machine language of the PDP-8 computer. Since the compilation consists in its majority of subroutine calls (to be compiled, after removal of LISP parentheses by MACRO-8) the technique is applicable with trivial modification to any other computer having the subroutine jump and indirect transfer instructions. Ile purpose of the program is both to compile REC expressions and to illustrate the workings of the REC language, and accordingly a description of this language is given. It contains operators and predicates; flow of control is achieved by parentheses which define subexpressions, colon which implies iteration, and semicolon which terminates the execution of an expression. Predicates pass control to the position following the next colon or semicolon, allowing the execution of alternative expression strings.